## 1.1    The Development of the Linac Coherent Light Source RF Gun

David H. Dowell, Erik Jongewaard, James Lewandowski,
Cecile Limborg-Deprey, Zenghai Li, John Schmerge, Arnold Vlieks,
Juwen Wang and Liling Xiao

Mail to: dowell@slac.stanford.edu
Stanford Linear Accelerator Center, Menlo Park, CA, US

### 1.1.1    Introduction

The Linac Coherent Light Source (LCLS) is the first x-ray laser user facility based upon a free electron laser (FEL) requiring extraordinary beam quality to saturate at 1.5 angstroms within a 100 meter undulator.[1]  This new type of light source is using the last kilometer of the three kilometer linac at SLAC to accelerate the beam to an energy as high as 13.6 GeV and required a new electron gun and injector to produce a very bright beam for acceleration.  At the outset of the project it was recognized that existing RF guns had the potential to produce the desired beam but none had demonstrated it. Therefore a new RF gun or at least the modification of an existing gun was necessary.

The parameters listed in Table 1 illustrate the unique characteristics of LCLS which drive the requirements for the electron gun as given in Table 2.  The gun beam quality needs to accommodate emittance growth as the beam is travels through approximately one kilometer of linac and two bunch compressors before reaching the undulator.

**Table 1:** Specifications of the Linac Coherent Light Source.

| Parameter | Value | Value | Units |
|---|---|---|---|
| Fundamental FEL Wavelength | 1.5 | 15 | Angstroms |
| Electron Beam Energy | 13.6 | 4.3 | GeV |
| Normalized Slice Emittance | 1.2 | 1.2 | microns (rms) |
| Peak Current | 3.4 | 3.4 | kA |
| Energy Spread (slice rms) | 0.01 | 0.03 | % |
| Bunch/Pulse Length (FWHM) | ~200 | ~200 | fs |
| Saturation Length | 87 | 25 | m |
| FEL Fundamental Power @ Saturation | 8 | 17 | GW |
| FEL Photons per Pulse | 1 | 29 | $10^{12}$ |
| Peak Brightness @ Undulator Exit* | 0.8 | 0.06 | $10^{33}$ |

*photons/sec/mm²/mrad²/0.1%-BW*

These beam requirements were demonstrated during the recent commissioning runs of the LCLS injector and linac [2] due to the successful design, fabrication, testing and operation of the LCLS gun.  The goal of this paper is to relate the technical background of how the gun was able to achieve and in some cases exceed these requirements by understanding and correcting the deficiencies of the prototype s-band RF photocathode gun, the BNL/SLAC/UCLA Gun III.

**Table 2:** Requirements of the LCLS gun

| Parameter | Value |
|---|---|
| Peak Current | 100 A |
| Charge | 1 nC |
| Normalized Transverse Emittance: Projected/Slice | <1.2 / 1.0 micron (rms) |
| Uncorrelated Energy Spread | 3keV (rms) |
| Repetition Rate | 120 Hz |
| Peak Cathode Field | 120MV/m |
| Gun Laser Stability | <0.50 ps (rms) |
| Gun RF Phase Stability | <0.1 deg (rms) |
| Quantum Efficiency | 3x10$^{-5}$ |
| Charge Stability | <2 % (rms) |
| Bunch Length Stability | <5 % (rms) |

This paper begins with a brief history and technical description of Gun III and the Gun Test Facility (GTF) at SLAC, and studies of the gun's RF and emittance compensation solenoid. The work at the GTF identified the gun and solenoid deficiencies, and helped to define the specifications for the LCLS gun. Section 1.1.5 describes the modeling used to compute and correct the gun RF fields and Section 1.1.6 describes the use of these fields in the electron beam simulations. The magnetic design and measurements of the emittance compensation solenoid are discussed in Section 1.1.7. The novel feature of the LCLS solenoid is the embedded quadrupole correctors. The thermo-mechanical engineering of the LCLS gun is discussed in Section 1.1.8, and the cold and hot RF tests are described in Section 1.1.9. The results of this work are summarized and concluding remarks are given in Section 1.1.10.

### 1.1.2 The History and Characteristics of the BNL/SLAC/UCLA S-Band Gun III and the Gun Test Facility (GTF) at SLAC

The SLAC Gun Test Facility (GTF) was initiated by Herman Winick in the early 1990s to test and develop the electron source necessary to drive the Linac Coherent Light Source. Following the proposal [3] for a short wavelength SASE FEL using the SLAC linac, a study group led by Pellegrini and Winick investigated the requirements. Foremost among these was a high brightness electron gun with performance beyond that of any in existence at that time. This led to the GTF project to develop an appropriate laser-driven, high gradient RF photoinjector.

The GTF was located in the SSRL injector linac vault taking advantage of the existing RF power, radiation shielding, safety interlocks, and staff technical support. The accelerator (a 3m S-band linac section made available by SLAC) and diagnostics were installed and commissioned under the leadership of John Schmerge and the late James Weaver. The laser was developed by graduate student David Reis in collaboration with David Meyerhofer, both from the University of Rochester. The first gun to be tested was called the Next Generation Photoinjector, a design developed largely by Stanford graduate student Dennis Palmer under the supervision of Roger Miller [4]. Completing the design and fabricating four copies of this gun was a collaborative effort involving BNL, SLAC and UCLA. This 1.6 S-band cell gun, hereafter referred to as Gun III, incorporated two significant improvements over the existing photocathode RF guns:

1. The port through which the RF entered the gun was quasi-symmetrized by adding an identical port on the opposite side which is connected to a vacuum pump. Since the opposing port is not an RF power feed it can only symmetrize the amplitude and not the phase. Due to the power flow through the single port, there remains a phase shift transverse to the beam in the full cell. This type of symmetric RF feed reduces but does not eliminate the dipole mode introduced by a single RF feed.

2. The gun and RF drive system were designed for higher gradient operation, since simulations indicated lower emittance at high charge was achievable at field levels around 140 MV/m.

Four of these guns were machined at UCLA and then brazed and cold tested at SLAC. High power testing proceeded at the GTF in 1996 and 1997. Two of the guns were used at BNL (one at the ATF and the other at the BNL Source Development Laboratory, SDL), a third went to UCLA and the fourth was characterized in detail at the GTF.

First beam at the GTF occurred in summer 1997 [5]. Subsequent measurements demonstrated the emittance dependence on laser pulse length [6]. Two PhD theses were completed at the GTF [7]. In the first years beam brightness was primarily limited by cathode uniformity. Improved diagnostic techniques let to improved beam brightness.

The UCLA and GTF guns differed slightly from the two BNL guns. The BNL guns used a helico flex seal to make both the vacuum and RF seal between the gun body and the cathode plate. The UCLA and GTF version shown in Figure 1 use a conflat flange behind the cathode plate for the vacuum seal and the RF seal is a press fit joint. Like most S-band guns, it uses a metal cathode which is incorporated into the rear wall of the gun. If the cathode material is copper, then the electrons are simply photo-emitted from the center of the approximately 10 cm diameter cathode plate which makes RF contact with the gun body at its outer diameter. In the UCLA and GTF design, the entire cathode plate is inside the vacuum envelope of the gun. Thus it is necessary to vacuum pump the volume behind the cathode plate simultaneously with the gun's RF volume, or to place a ring of through holes near the outer diameter to allow the gases in the volume behind the cathode to vent into the gun. Several holes are required to give sufficient pumping.

The RF power is coupled into the gun through a hole in the full cell and power flows into the half cell through the beam iris. An un-powered hole opposite the power coupled port helps to reduce the dipole RF field and is used for vacuum pumping. The un-powered port also has an RF pickup probe. These kidney shaped ports have their long axes aligned in the azimuth direction to minimize the quadrupole RF field.

At the center of the cathode plate rear is a brazed stainless steel insert and nut. Not shown in the drawing is the bellows and differential screw which are used to slightly deflect the center into or out of the cathode cell while the gun is under vacuum. The deflection changes the gun volume to allow adjustment of the gun's resonant frequency.

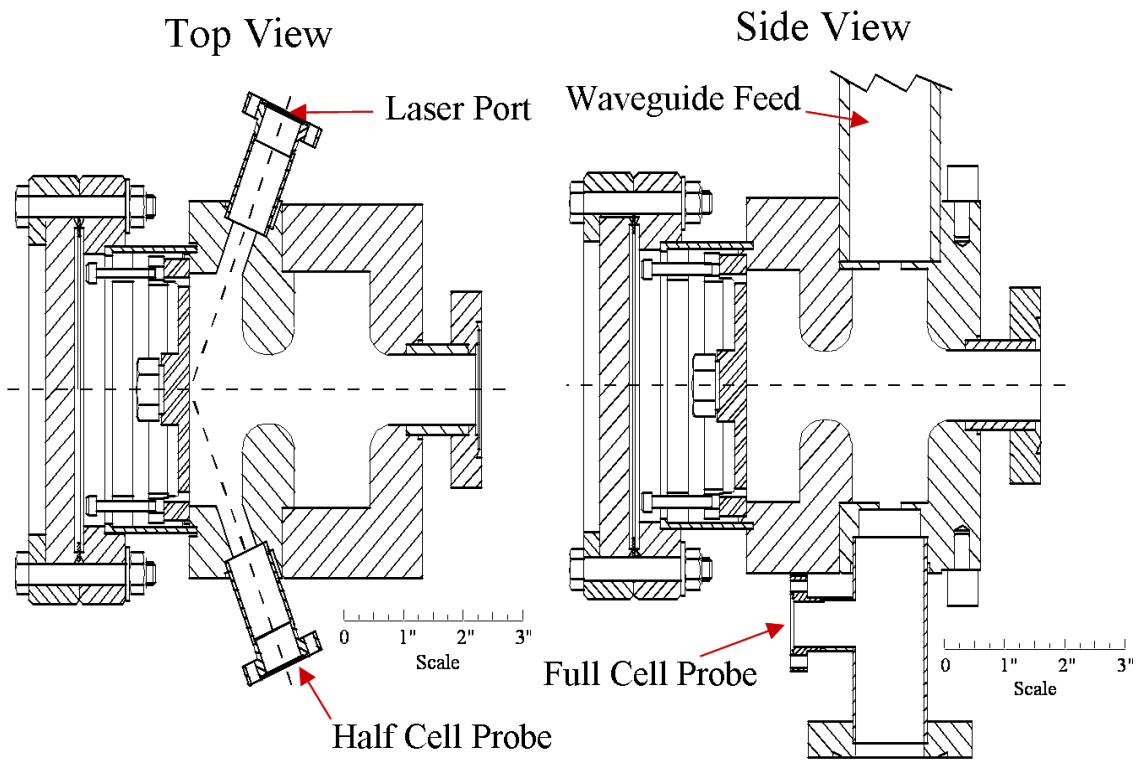

**Figure 1:** The BNL/SLAC/UCLA Gun III 1.6 cell S-band RF gun.

### 1.1.3 Experimental Studies of the GTF Gun III

Although the GTF Gun III did not produce a beam with an emittance low enough for LCLS, its construction and operation did provide valuable information which led to the LCLS gun which did achieve the stringent beam parameters. The GTF work identified two important technical problems which were solved in the LCLS gun design. The first was the gun produced a bunch with a large correlated energy spread and the second was the presence of quadrupole fields in the emittance compensation solenoid. The correlated energy spread was first observed during studies of the longitudinal phase space and results from the RF excitation of the zero-mode along with the π-mode in the two cell gun. This occurs because the frequency separation of these two modes is only 3.5 MHz so the tail of the 0-mode extends into the π-mode resonance allowing the 0-mode to be excited as well.

The solenoid field was carefully studied after observing an asymmetry in the beam profiles. Detailed magnetic measurements showed quadrupole fields at the solenoid's entrance and exit. While these field errors were small, it was decided to correct for them with weak normal and skew quadrupole correctors which had additional benefits discussed below.

The results of the GTF work related to the mode-beating and the solenoid quadrupole fields are described in some detail in the following sections.

### *1.1.3.1   Longitudinal Phase Space Studies*

Here we discuss only those aspects of the GTF beam experiments which are relevant to the design of the LCLS gun. Details of the GTF longitudinal phase space and slice emittance studies can be found in Ref [8].

The correlated energy spread first exhibited itself during our measurements of the longitudinal emittance and its phase space distribution. In these experiments the longitudinal emittance and Twiss parameters are determined from electron energy spectra after a 3-meter linac section as a function of the linac section RF phase. Data and simulations for 16 pC bunch charge are shown in Figure 2 of the rms bunch energy spread as functions of the linac phase. The measurements are at low charge in order to minimize the effects of space charge and longitudinal wakefields which can also increase the energy spread. The linac phase of zero degrees S-band (degS) is defined as maximum energy gain in the linac or on crest. The correlated energy spread from the gun can be estimated by the amount of chirp needed from the linac to produce the minimum energy spread. The final chirp to first-order is given by

$$\Delta E_1 = \Delta E_{gun} + \frac{dE}{d\phi}\bigg|_{linac} \Delta\phi_{linac}$$

The overall energy spread is a minimum when the linac introduces an energy chirp which cancels that coming from the gun,

$$\Delta E_{gun} = -\frac{dE}{d\phi}\bigg|_{linac} \Delta\phi_{linac} = E_{linac}\sin\phi_{linac}\Delta\phi_{linac},$$

where the linac energy gain is given by $E_{linac}\cos\phi_{linac}$. Using the observed 8 degS phase and the 30 MeV energy gain of the linac yields a correlated energy of 73 KeV/degS.

Also shown in Figure 2 are simulations of a particle-tracking code [9] which uses 0- and $\pi$-mode field distributions given by Superfish [10] to compute the energy spread after the linac. Unfortunately Superfish cannot give the phase relation between the two modes which comes from the RF dynamics while the gun is filling, therefore both the 0-mode field strength and the relative phase where varied to produce a best fit to the data. As expected the $\pi$-mode only simulation using 110 MV/m gives a minimum energy spread near 0 degS, but an additional 0-mode field of 13 MV/m with a 0-$\pi$ relative phase of -85 degS is needed to reproduce the data. [11]

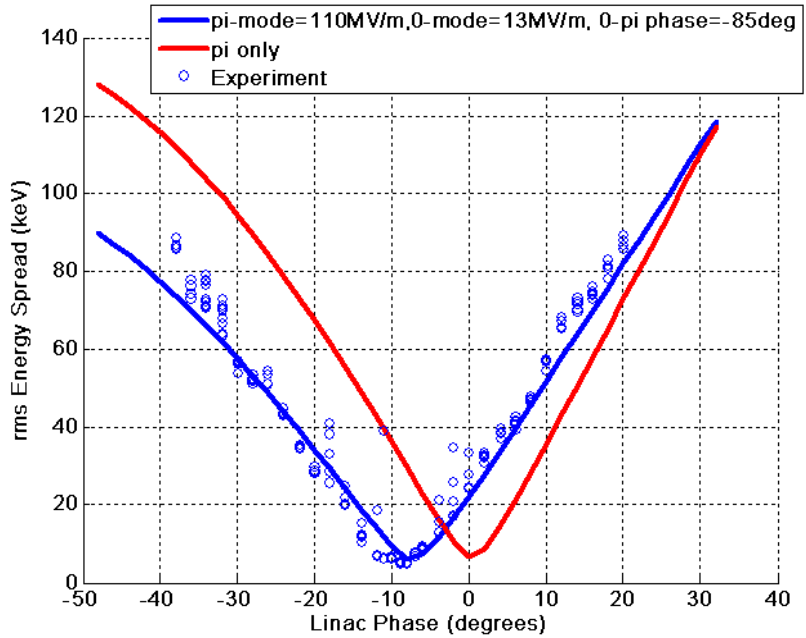

**Figure 2:** The measured and computed beam energy spread at 30 MeV as a function of the booster linac phase.

This analysis showed this energy spread results from the presence of the 0-mode which unbalances the total field between the two cells and chirps the beam. The 0-mode for the GTF Gun III is only 3.5 MHz lower than the desired $\pi$-mode and is easily excited. The next section describes RF measurement which verified the beating between the two modes at the mode separation frequency.

### 1.1.3.2 RF Measurements of 0-π Mode Beating

Given the large energy spread there were concerns about the physical condition of the GTF Gun III and whether or not it has somehow changed during operation. Therefore it was removed from the GTF beamline and its field balance measured using the bead drop (Slater perturbation) method. The results are given in Figure 3 for both the $\pi$- and 0-mode, along with a Superfish simulation [12]. The position of the cathode plate was adjusted in the simulation to match the $\pi$-mode data, and the 0-mode simulation was then done with the same parameters. The data show the gun was still balanced and was not the cause of the large correlated energy spread.

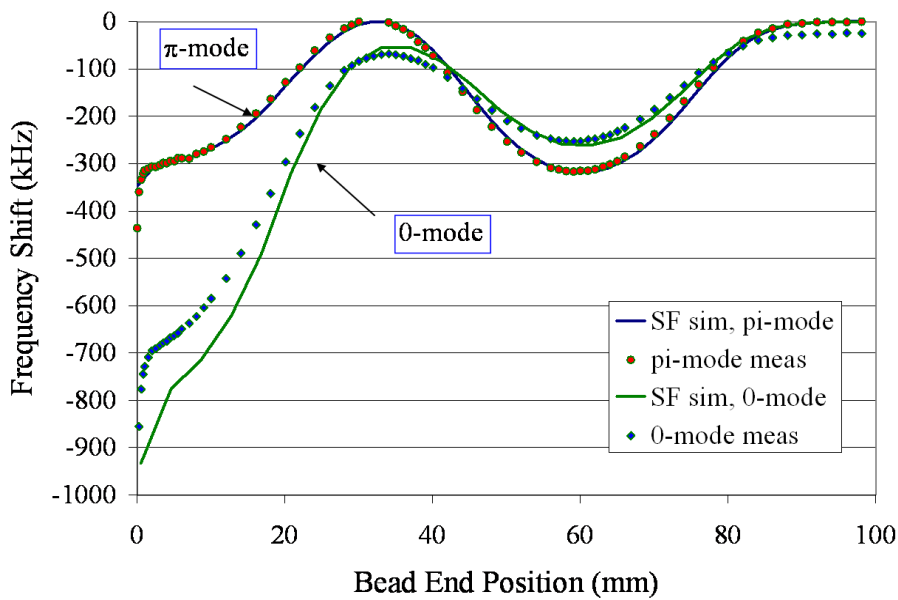

**Figure 3:** Beam drop measurement and simulation for the π- and 0-modes of the GTF gun.

In order to further investigate the cause of the energy spread, an additional RF probe was installed in one of the laser ports located on the cathode cell to compare with the full cell RF probe, and the gun re-installed on the beamline. Because of the π phase shift between the cells for the π-mode and no shift for the 0-mode, the difference of the two probe signals will exhibit an oscillation at the mode-spacing frequency if there is any 0-mode present. The results are shown in Figure 4 and clearly show the RF fields oscillating at the separation frequency of 3.5 MHz.

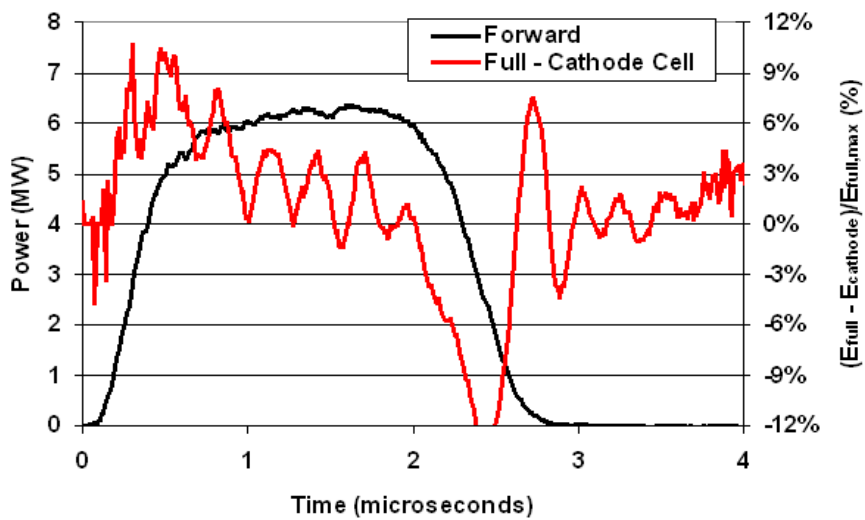

**Figure 4:** RF probe measurements of the GTF gun showing 0-π mode beating.

In conclusion, these RF studies demonstrate the correlated energy spread results from the presence of the 0-mode which unbalances the total field between the two cells and chirps the beam. The 0-mode for the GTF gun is 3.5 MHz lower than the π-mode and is easily excited by its overlap with the π-mode.

### 1.1.3.3    Studies of the GTF Gun Solenoid

In addition to the large correlated energy spread, there was also the observation of an asymmetric beam shape which motivated the investigation of the field quality of the gun solenoid, also known as the emittance compensation solenoid. Therefore the solenoid was removed from the beamline and installed on a measurement bench in the SLAC magnetic measurements lab. Here the solenoid field was thoroughly characterized using Hall probes and rotating coils. This experience was used to establish the characterization processes for the LCLS emittance compensation solenoids.

The short rotating coil data is plotted in Figure 5 for the dipole, quadrupole and sextupole fields of the GTF solenoid. The dipole field is due to a slight misalignment of the coil's axis of travel with the solenoid magnetic axis. (In fact, due to its sensitivity, the LCLS solenoid characterization procedure defines the magnetic axis as the line of zero dipole field as measured by the rotating coil.) The data in Figure 5 is taken along the magnetic axis with zero corresponding to the center of the solenoid which has an effective length of 19 cm. The quadrupole field distribution has a peak approximately 5 cm long at each end of the solenoid and, though the data is noisy) there is a 90 degree rotation between the two end fields.

An attempt was made to remove these quadrupole fields by re-designing and installing new coil windings, but without success. Therefore in the LCLS solenoid it was decided to cancel the fields with correctors forming normal and skew quadrupoles fields with eight single wires running the length of the solenoid's bore. This and other means for cancelling the quadrupole field error are described later.

The measured quadrupole fields are weak, having an equivalent focal length of approximately 20 meters and longer, which is weak compared to the solenoid's 12 cm focal length. However given the important role of the emittance compensation solenoid, and since the expected field strength was not understood or predictable, it was decided to include the corrector quadrupoles in the LCLS solenoid design. The implementation of the correctors and their unexpected benefit to the emittance are described in later sections.

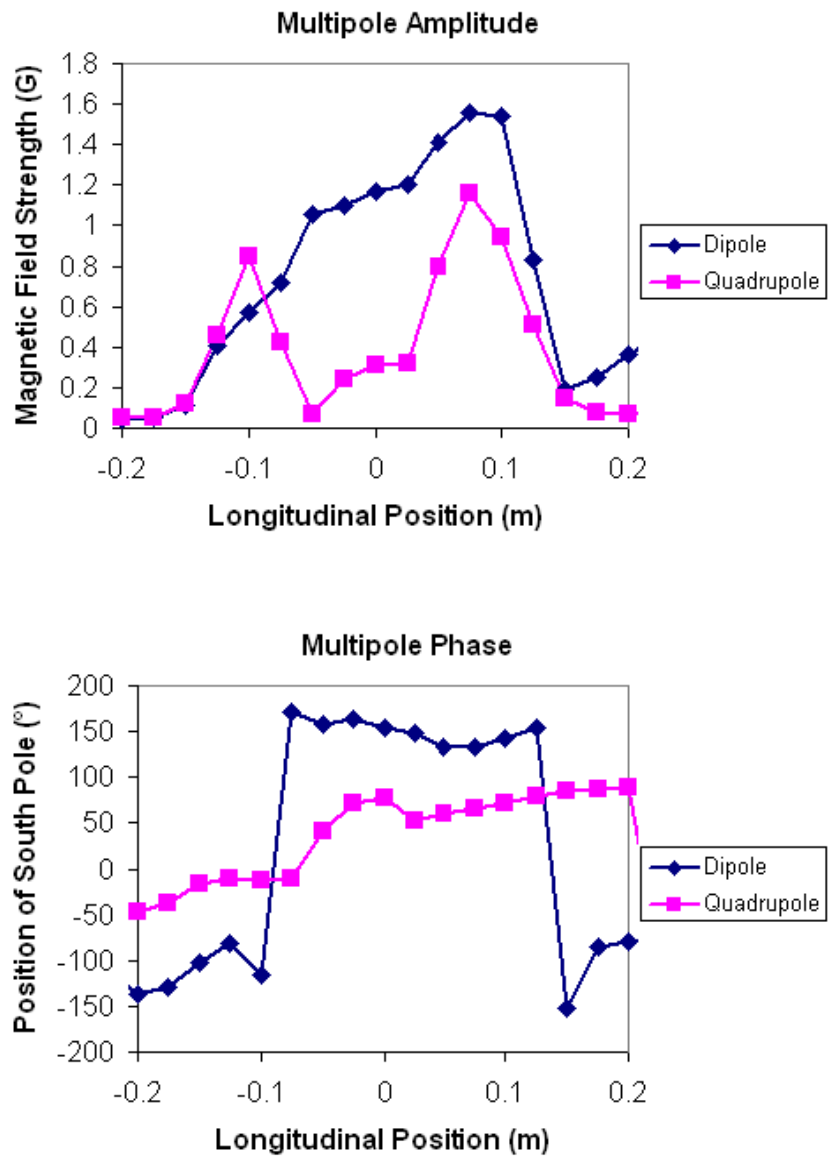

**Figure 5:** Rotating coil measurements of the GTF solenoid multipole fields. Top: The dipole and quadrupole field strengths at the 2.71 cm coil radius. Bottom: The phase angle for the two multipoles.

### 1.1.3.4    *Impact of GTF Work on the LCLS Gun Design*

Although the GTF did not achieve the beam quality needed for LCLS, it was instrumental in showing how to design and build a gun which did. As a result of this work and with the advice of the RF Gun Technical Review Committee [13], it was decided to increase the 0-π mode separation to 15 MHz and to include quadrupole correctors in the emittance compensation solenoid. Other design changes and improvements based upon the GTF experience include a modified cathode assembly to shorten the cathode replacement time and allow adjustment of the RF seal and RF resonance with the gun at UHV, and the elimination of slug tuners.

### 1.1.4   Final Design Characteristics of the LCLS Gun

In addition to the mode-beating and the solenoid quadrupole field errors, there were other technical problems to solve and incorporate into the new LCLS gun. These included pulsed RF heating which would reduce the gun's lifetime due to cracking of high-stress surfaces, RF breakdown on surfaces with high peak fields, wakefields between the gun exit and the booster linac entrance, RF field asymmetry in both the gun and booster linac, phase balance of the two arms of the dual RF feed, etc. These and other design and fabrication issues benefited from the guidance of the RF Gun Technical Review Committee, and many of their suggestions were incorporated into LCLS gun. [14]

A summary of changes made to Gun III which enabled the LCLS gun meet the stringent requirements of the x-ray FEL are:

1. Dual RF feed to eliminate any transverse RF field asymmetry due to the flow of RF power.
2. A racetrack shape in the full cell to correct for quadrupole fields introduced by the dual feed and other penetrations.
3. Increased the mode separation from 3 to 15 MHz which greatly reduces beating between the 0 and π RF modes.
4. The iris between the two cells was reshaped to reduce its surface field to be lower than the cathode field. This and the larger iris diameter needed to increase the mode separation had the added benefit of reducing the spherical aberration at the iris.
5. The RF power is coupled into the full cell using two longitudinal rectangular ports running the full length of the cell. The combination of z-coupling and increasing the radius of the edges greatly reduces the pulsed heating.
6. Deformation tuners consisting of studs brazed into areas where the walls are thinned allow for small tuning corrections to the resonant frequency and cell-to-cell field balance. The slug tuners used on the GTF Gun III were a source of field breakdown and mechanical failure. The LCLS gun deformation tuners were never used as the machining was done within 0.0004 inches of the design dimensions.
7. Cooling channels capable of dissipating 4 KW of average RF power. At 120 Hz this corresponds to a cathode peak field of 140 MV/m.
8. The cathode was designed for rapid replacement with a new mounting allowing for adjustment of the RF seal and resonance frequency with the gun under vacuum. This design is compatible with a future load lock for installing cathodes needing UHV.
9. Dipole and quadrupole field correctors incorporated into the magnetic solenoid. The gun solenoid was fully characterized using Hall probe and rotating coil measurements.
10. A bucking coil was added to cancel the small magnetic field at the cathode and the emittance growth it causes.

The following subsections describe how these features were incorporated into the RF design of the LCLS gun and the emittance compensation coil. The thermo-mechanical engineering details are given in Section 1.1.8.

### 1.1.5   RF Modeling and Design

#### 1.1.5.1   Numerical Simulations

The parallel eigensolver Omega3P was used to design the LCLS RF gun [15]. Omega3P is one of codes in the high performance computing electromagnetic tools developed at SLAC. It is based on finite-element unstructured meshes and parallel computation implementations on supercomputers, and is capable of simulating large complex RF systems with unprecedented resolution and turnaround time. It has been successfully applied to many existing and future accelerator R&D projects to improve the machine performance and to optimize the designs [16].

#### 1.1.5.2   1.6-cell gun 2D shape optimization

The LCLS RF gun operates in the $\pi$ mode with f = 2.856 GHz. The gun also supports a 0-mode that is below the operating mode frequency. In the standard GTF RF Gun III, the mode separation between the 0- and $\pi$-mode is 3.4 MHz. Because of the finite bandwidth, the amplitude of 0-mode in the half cell is about 10% that of the $\pi$-mode when steady state is reached [13, 17], which was found having significant effects on the beam emittance. In the LCLS RF Gun, this mode separation is increased to 15 MHz to reduce the 0-mode excitation in the half-cell to less than 3%. The 15 MHz mode separation was achieved by increasing the aperture of the iris between the full and half cells and by reducing the disk thickness. In addition, modifying the disk and cutoff hole shapes from circular to elliptical can reduce the peak surface field there from 11% higher to 2% lower than the field on the cathode with the shunt impedance maintained at the same value. A 2D computer model of 1.6 cell RF gun with the modifications described is shown in Figure 6.

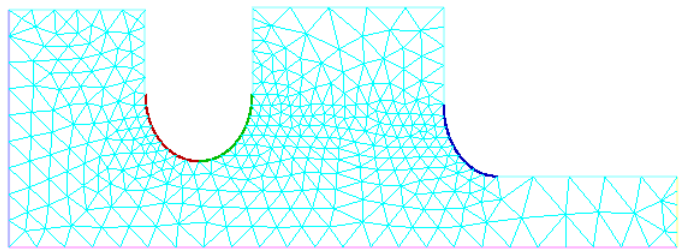

**Figure 6:** The 2D mesh of the LCLS gun cavity as modeled by Omega2P.

#### 1.1.5.3   1.6-cell gun coupler design

Based on the 2D cavity shape, Omega3P was used to model the 3D gun structure that includes the input couplers as well as the laser ports. The LCLS RF gun adopted a dual-feed scheme. A quarter model of the RF gun with the input coupler is shown in Figure 7. The boundary condition at the end of the waveguide was set to be matched. The complex eigensolver in Omega3p then was able to calculate the resonant frequency $f$, quality factor $Q_0$ and the external $Q_{ext}$. The design requirement for the coupling beta for the LCLS RF gun is 2. The coupler iris is rounded to minimize the RF heating. The dual feed couplers break the azimuth symmetry and induce quadrupole field components in the full cell. These are compensated with a race track interior shape for the full cell.

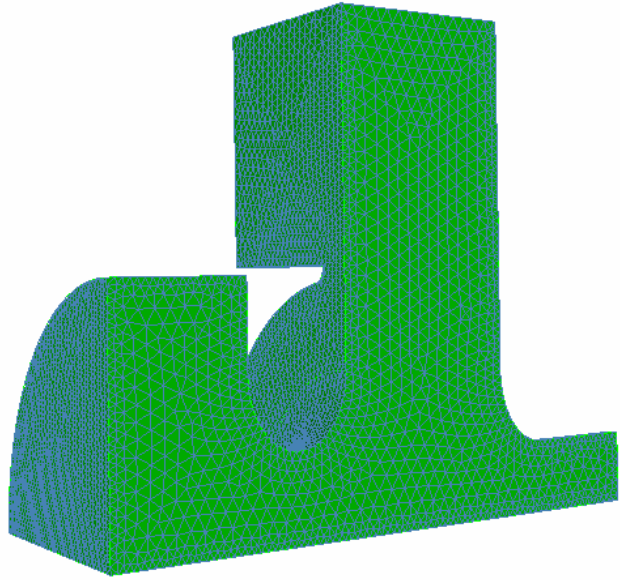

**Figure 7:** A quarter model of the LCLS RF gun interior surface.

### 1.1.5.4 *Pulsed Heating*

In the single feed design of Gun III, θ-coupling was adopted to reduce the dipole field. At 120Hz operation for this design, the temperature rise at the end of the coupling aperture where it is curved could reach 150° C. This pulsed heating will seriously limit the gun's life time [18]. Based on the NLC experience, the temperature rise due to the pulsed heating should be below 50 °C. A straightforward way to reduce the heating is to increase the radius on the inside surface of the coupler aperture. However, this rounding of the radius is difficult to machine with θ-coupling so z-coupling is used instead. With z-coupling, because the iris has straight sides which extend the full length of the cell, the required radius can easily be fabricated.

In the z-coupling scheme, the width of the coupling slot and the rounding radius were adjusted to obtain a coupling coefficient around 2 and a temperature rise below 50° C. The following equations were used to evaluate the temperature rise at the end of a RF pulse [19].

$$\Delta T_{max} = \frac{R_s}{K} \sqrt{\frac{D}{\pi}} \frac{1}{2} \int_0^{t_p} |H_{s\,max}(t)|^2 \frac{dt}{\sqrt{t_p - t}}$$

$$R_s = \sqrt{\frac{\omega\mu}{2\sigma}} = \frac{1}{\sigma\delta_s}, \qquad \delta_s = \sqrt{\frac{2}{\omega\mu\sigma}}$$

Where $K=360W/m/^0C$ is the thermal conductivity, $D=1.132 \text{x} 10^{-4} m^2/sec$ is the specific heat of copper. Here $H_{smax}$ refers to the maximum surface magnetic field along the coupling slot on the inside of the cell and will decrease as the rounding radius $r2$ increases. Figure 8 shows the results calculated assuming that the maximum electric field on the cathode is 120MV/m, the coupling coefficient is 2 and the pulse length $t_p$ is 3μs. A 1.5mm rounding (r2) on the iris will reduce the RF heating to below $50^0$C. However a thicker disk is required in order to reduce the thermal stress. In the final design, the iris rounding $r1$ and $r2$ were determined to be 1.0mm and 7.5mm respectively, and the width of the z-coupling slot to be 16.5 mm.

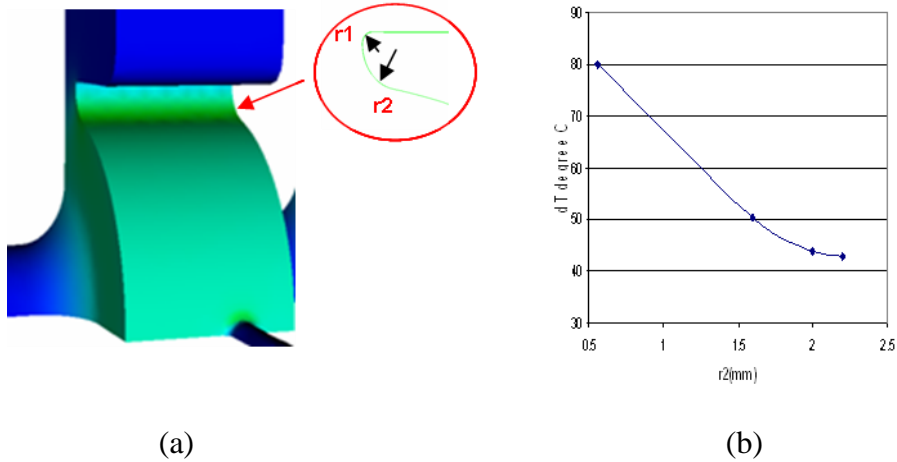

(a)                                                          (b)

**Figure 8:** (a) Surface magnetic field distribution;

(b) Temperature rise vs. rounding radius r2.

### *1.1.5.5     Quadrupole RF Fields*

While the dipole field is removed by the dual feed design, the quadrupole component remains unaffected so a racetrack shape has been adopted for the coupler cell to reduce its effect as shown in Figure 9a. In this geometry, the center offsets of the two circles were adjusted to minimize the quadrupole field on the beam axis. This has led to a reduction of the maximum quadrupole moment γβr from $4.4 \text{x} 10^{-3}$/mm to $8 \text{x} 10^{-5}$ /mm. Field determination to this level of accuracy was only possible by using $4^{th}$ order basis functions in Omega3P. Field maps generated with Omega3P were used in beam dynamics calculations of the gun emittance [17] and found significant improvements in beam emittance over the un-compensated design.

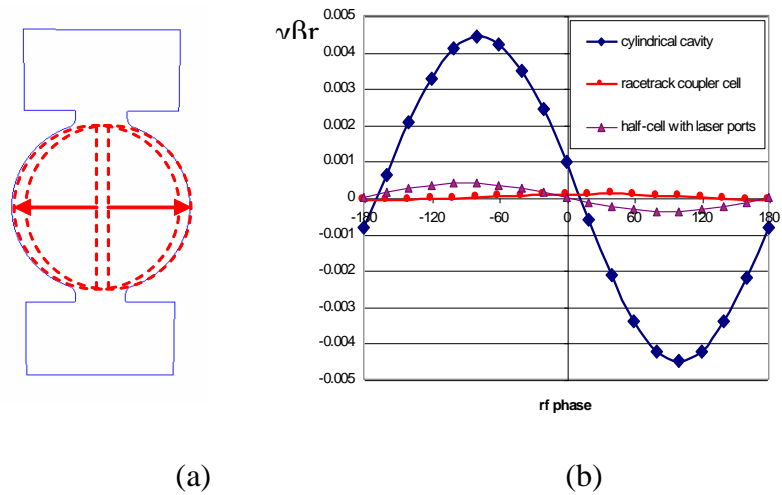

<table>
<tr><td>(a)</td><td>(b)</td></tr>
</table>

**Figure 9:** (a) Racetrack coupler cell; (b) Quadrupole moment in the gun cavity [15]

### *1.1.5.6    Laser and RF Probe Ports*

Figure 10 shows the details of the laser and RF probe ports. The two laser ports in the half cell are on the horizontal plane (shown here in the vertical plane) and admit the laser beam through an elliptical aperture. The effects of the laser ports on the frequency and field balance were compensated by adjusting the half-cell radius. The quadrupole moment in the half cell introduced by the laser ports was found to be about $\Delta(\gamma\beta\perp) = 3.85\times10^{-4}$/mm (see Figure 9(b)) which, as shown from PARMELA simulations, slightly changed the tuning but did not degrade either the slice or the projected emittance [17]. There are two pickup probe ports in each of the full and half cells. They are azimuthally 90 degrees from the power couplers and the laser ports. Both the laser and pickup ports are rounded with a 1.5-mm radius to minimize the RF pulse heating. The maximum RF heating temperature rise at the ports is less than 30°C.

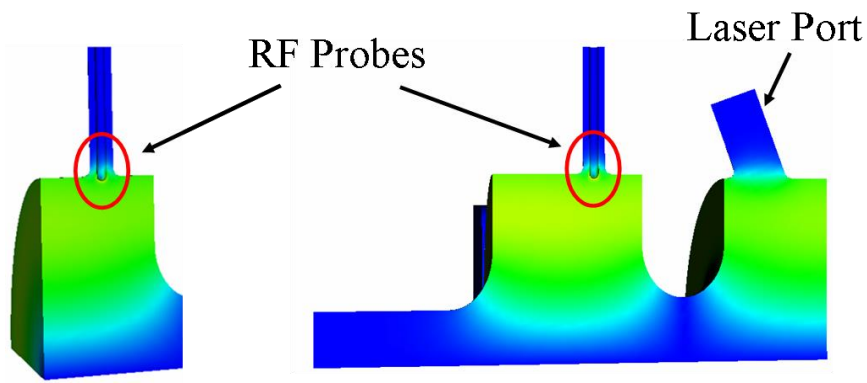

**Figure 10:** One quarter model showing the RF probe ports for the cathode cell (left) and the full cell (right). The laser port on the cathode cell is at a 22.5 degree angle relative to the cathode surface.

The pickup probes are calibrated to monitor the field balance as well as the field gradient in the gun cavities. A -50dB coupling is obtainable with the probe tip flush with the cavity radius. Considering the requirements in power handling of the electronics and accurate measurement of the fields, a coupling around -60dB coupling is needed, which requires that the probe tip be slightly recessed behind the cavity radius. The probe was initially made of stainless steel. The RF heating on the probe tip was evaluated to be as much as 100 °C at the nominal operating cathode field and pulse length. During the high power processing of the LCLS Gun1, it was found that this heating can cause significant change in coupling, causing inaccurate readings in cavity gradient and excessive vacuum load. Two improvements were made to mitigate this problem: 1) the coupling was lowered to -77 dB, the tip of the probe retreats more behind the cavity radius which minimizes RF heating; 2) the probe was copper plated, further reducing the RF heating by a factor of 7.

### 1.1.5.7    *The Final LCLS Gun RF Design and Comparison with Gun III*

The RF design described above produced the final shape of the interior surfaces of the LCLS gun which could be used to produce the engineering design. The left of Figure 11 shows the solid model of the gun's interior surfaces, illustrating the z-coupling of the vertical waveguide feeds to the full cell, and the horizontal location of one of the two RF probes. The half cell portion of the solid model shows one of the laser ports in the horizontal plane and its two RF probes positioned vertically. The geometry of the full cell racetrack shape is produced by two circles offset horizontally by 5.9 mm as illustrated in the left drawing of Figure 11 [15].

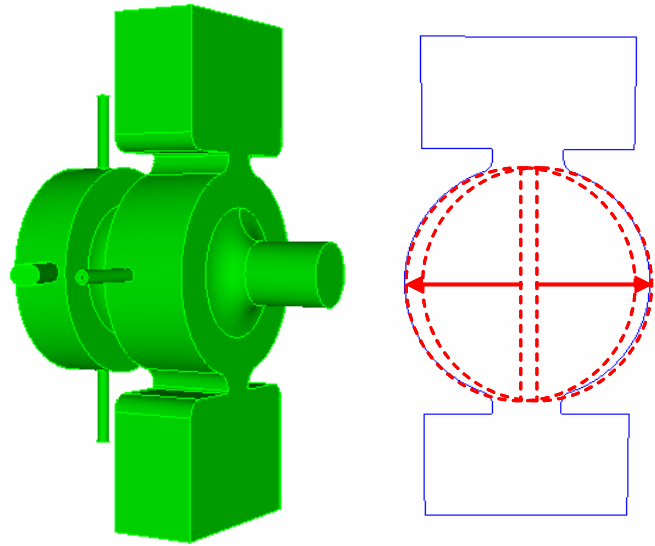

**Figure 11:** Interior geometry of the LCLS gun. Left: The solid model of the interior surfaces. Right: Drawing of the full cell showing the construction of the racetrack shape with two offset circles.

A comparison of the parameters for the BNL/SLAC/UCLA Gun III and the LCLS Gun is given in Table 3.

**Table 3:** The BNL/SLAC/UCLA Gun III and the LCLS Gun parameters

| | BNL/SLAC/UCLA Gun III | LCLS Gun |
|---|---|---|
| cathode field | 120MV/m | 140MV/m |
| rf feed | single w/compensation port | dual feed |
| cavity shape | circular | racetrack |
| 0-π mode separation | 3.4MHz | 15MHz |
| repetition rate | 10Hz | 120Hz |
| peak quadrupole field | 4 mrad/mm | 0.1 mrad/mm |
| RF tuners | plunger/stub | deformation |
| cathode | copper or Mg | copper |
| rf coupling | theta (azimuth) | z (longitudinal) |
| β-coupling | 1.3 | 2.0 |
| laser incidence | grazing or normal | grazing or normal |

### 1.1.6 Electron Beam Simulations

Simulations of the gun were performed to evaluate the effect of the mode separation on the beam quality using the two-frequency version of Parmela developed for studies of the two frequency gun [20]. Calculations were done for the nominal LCLS case of 1 nC, 1.2 mm radius laser on the cathode and 0.72 micron thermal emittance. The emittance was evaluated at 135 MeV after the two accelerator sections in the LCLS injector configuration where the emittance has damped to its final value. Since it is uncertain without knowing the details of the RF driving the gun, the simulations were done as a function of the phase between the two modes. The studies are summarized by four cases shown in Figure 12. The horizontal lines give the emittances obtained after optimizing with the pure π-mode fields of the gun geometries corresponding to 3.5 and 15 MHz mode separations. The two curves show the emittance as a function of the phase between the two modes. The amplitudes used for the 0-mode are 12 MV/m and 3 MV/m for 3.5 MHz and 15 MHz, respectively. These amplitudes are based on time-dependent RF field calculations of coupled RF resonances driven by a square pulse. The 12 MV/m amplitude used for 3.5 MHz separation is consistent with the 0-mode amplitude obtained from the fits to GTF energy spread data shown in Figure 2. Parameters such as the emittance compensation solenoid field have been adjusted to minimize the projected emittance for a 0-π relative phase of 90 degS.[17]

As expected the emittances for 15 MHz separation are nearly insensitive to the phase difference over a large range in comparison to the strong dependence of the 3.5 MHz separation. In addition, the simulations also show an overall lower emittance for the larger separation even when there's only π-mode field present. While not verified, it is thought that this lower emittance results from a reduction in the spherical aberration produced by the iris between the two cells due to the increased iris diameter needed to produce the 15 MHz gun geometry.

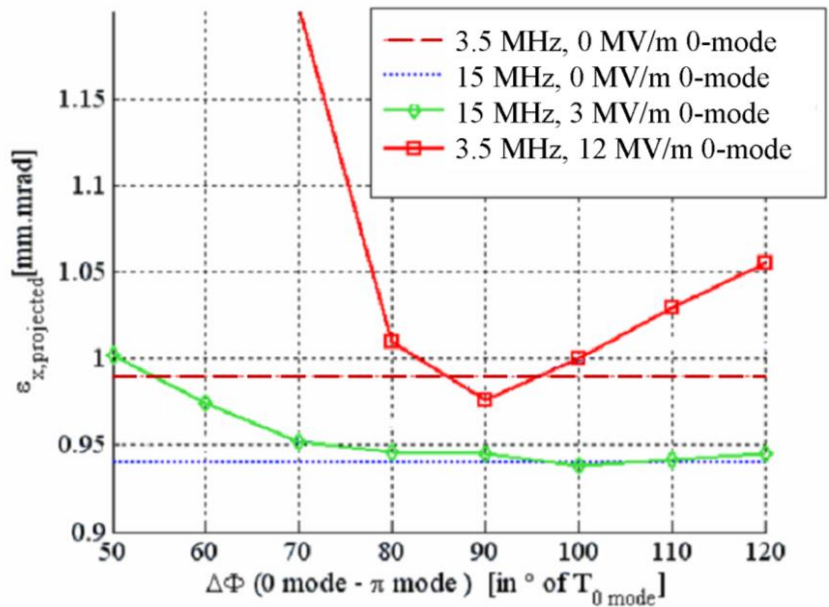

**Figure 12:** The projected emittance as a function of the phase difference between the 0- and π-modes for 3.5 MHz and 15 MHz separation frequency [17].

Other benefits of the increased mode separation which were noted during these and other studies are reduced sensitivity of the field balance to the gun body temperature and the gun's natural selection of only the upper branch of the field balance tuning curve (see Section 1.1.9.1).

### 1.1.7   The Emittance Compensation Solenoid

The LCLS emittance compensation solenoid, aka the gun solenoid, is similar to that used with the GTF gun, with the principle differences being a slight thinning of the end plates to move the solenoid closer to the gun and the addition of quadrupole correctors. Figure 13 is a photograph of the solenoid mounted on a test stand at the SLAC magnetic measurements lab. The water lines can be seen coming out of the solenoid's top, and the power cables are twisted to the left. The black coil mounted at the rear is the bucking coil which cancels the solenoid's cathode magnetic field. The bucking coil is being held in the proper location relative to the solenoid for magnetic characterization. In the lower, right foreground is the armature for the rotating coil used to determine the magnetic multipole fields.

The solenoid was magnetically characterized with a Hall probe to determine the effective length and its excitation calibration, and then with a 2.5 cm long, 2.71 cm radius rotating pickup coil to determine the longitudinal dependence of the magnetic multipoles. The quadrupole field measurements as a function of position along the solenoid's centerline are plotted in Figure 14. Similar to the GTF solenoid data shown earlier, the quadrupole field has peaks at each end of the magnet with a relative rotational phase shift of 90 degrees.

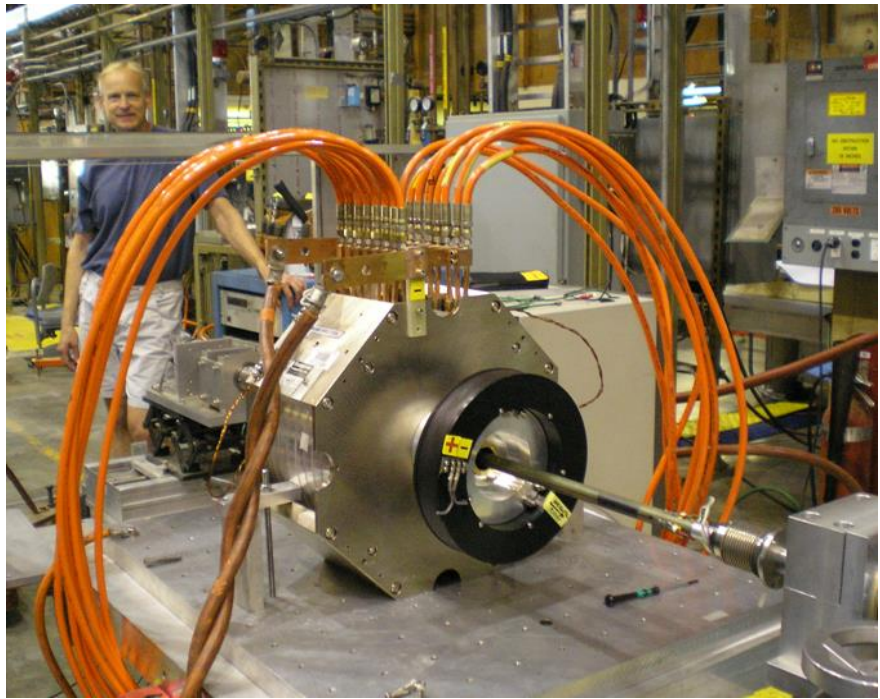

**Figure 13:** The emittance compensation solenoid and bucking coil in the SLAC magnetic measurements lab.

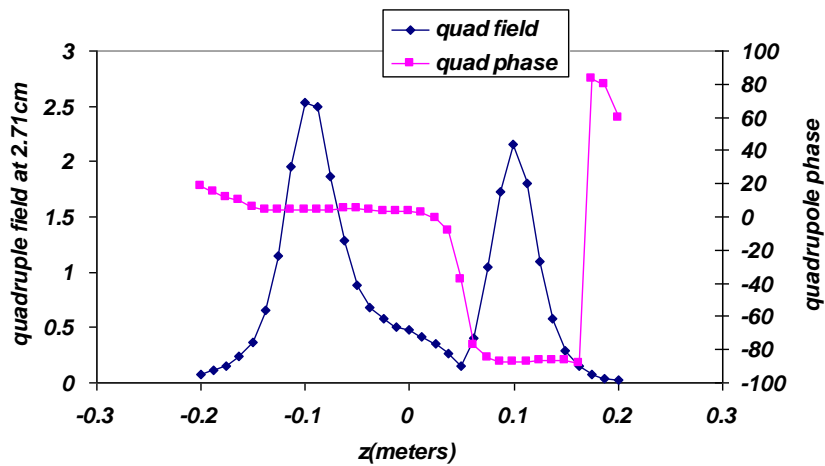

**Figure 14:** Rotating coil measurements of the quadrupole field and phase along the solenoid's magnetic axis.

Two gun and solenoid assemblies defined by the UHV envelope have been built for LCLS. While the guns are identical, the solenoids differ in the means used to cancel the end quadrupole fields. In the Gun1 solenoid, the quadrupole correctors consist of eight, single 12 AWG wires running the length of the solenoid bore arranged as normal and skewed quadrupoles. This configuration corrects for the quadrupole error averaged over the full length of the solenoid. It is relevant to note that at the nominal solenoid field for emittance compensation, the beam rotates approximately 110 degrees in the lab frame between the entrance and exit of the solenoid, which nearly matches the relative 90 degree rotation between the quadrupole end fields. In any case, given the delicate nature of the emittance compensation, the solenoid field quality should and could be improved by local cancellation of the field errors at the ends.

Therefore short, printed circuit board quadrupoles, PC quads, were installed at each end of the solenoid for Gun2. The PC quads were kindly provided by University of Maryland where they are used in the UMER low energy electron ring. As the diameter of the UMER PC quads was too small, the poles were cut apart and into the inner diameter of the solenoid's bore, as shown in the Figure 15 photographs. Unfortunately there was insufficient space to install normal and skew PC quads at each end, therefore a single PC quad was installed at each end, and rotated to be aligned with the orientation solenoid's end field. In addition, the PC quads could not extend past the ends of the solenoid otherwise they would interfere with the gun at one end and vacuum valve at the other.

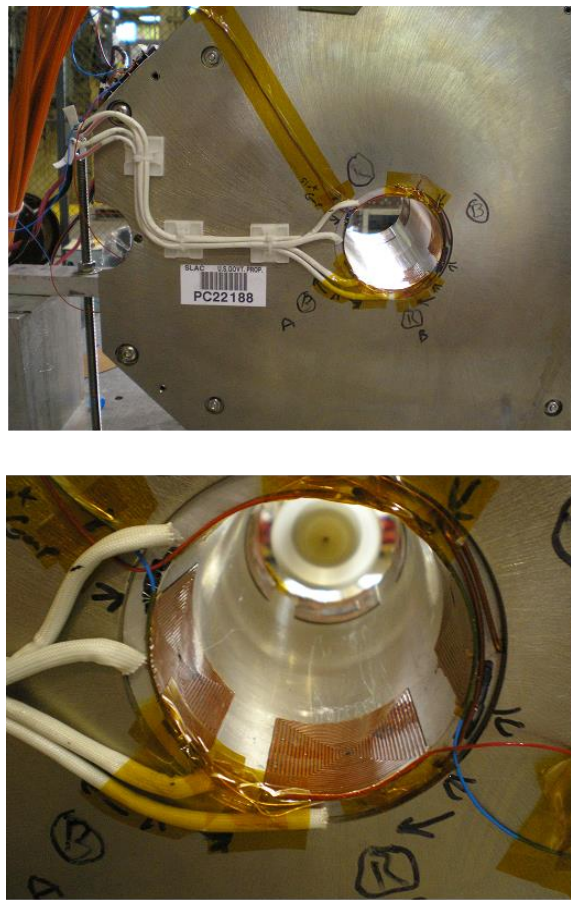

**Figure 15:** Photographs of the solenoid for Gun2 showing the single wire and PC quadrupole correctors. The rotating coil used for measuring the field multipoles can be seen at the far end of the solenoid bore.

The quadrupole field distribution measured along the axis of a PC quad is plotted in Figure 16 which when combined with the solenoid quadrupole field is a little too short. Figure 17 illustrates this for a PC quad centered at the peak and rotated to the angle of the solenoid field to cancel the total quadrupole field at the z = -0.1 meter end of the solenoid. With a slightly longer PC quad the cancellation could be made exact.

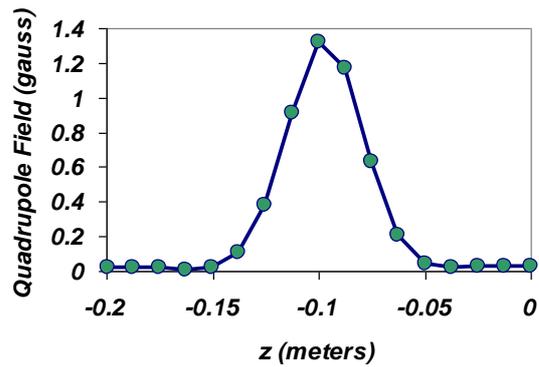

**Figure 16:** Axial field distribution for a PC quadrupole corrector.

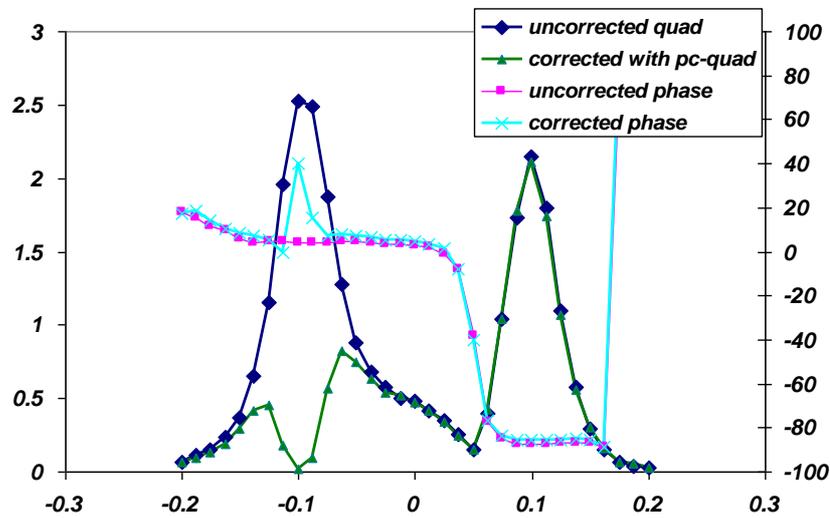

**Figure 17:** The uncorrected and corrected quadrupole field strength and phase for a PC quadrupole optimally placed at the peak of the uncorrected quadrupole field.

Although quite good cancellation is obtained in Figure 17, as described above, in this position the PC quads interfere with the gun and vacuum valve since they extended proud of the physical ends of the solenoid. Therefore it was necessary to mount them flush with the ends and thus extend too far inside the solenoid. The result of this unfortunate necessity is shown in Figure 18, in this case for the optimum cancellation with PC quads at both end of the solenoid. The longitudinally shifted PC quad results in a bipolar field distribution. The quality of the data can be seen by the excellent agreement between the calculated difference of the measured solenoid field and the PC quad fields with measurements of the cancelled fields.

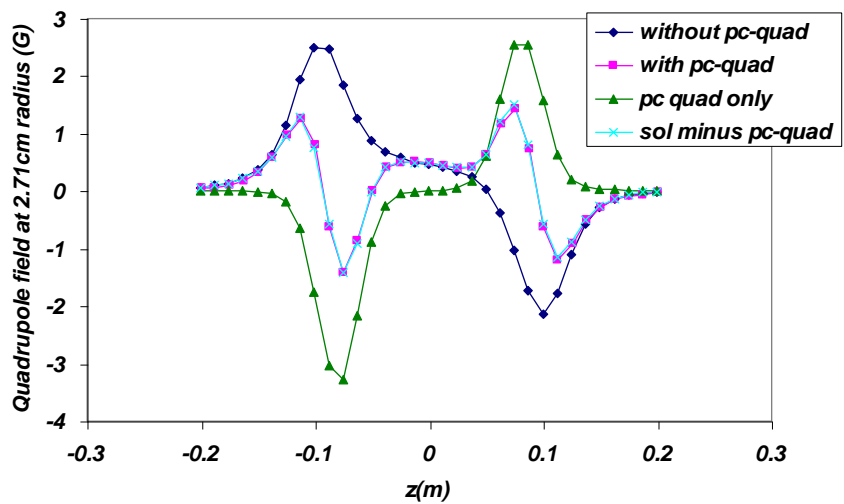

**Figure 18:** Measurements of the quadrupole magnetic field through the solenoid for the solenoid alone, the PC quads alone, and with the PC quads optimized to cancel the local average field strength. The difference between the measured solenoid and PC quad field is also plotted.

To summarize, magnetic measurements of the LCLS solenoids show the same end quadrupole fields as first found for the GTF solenoid. In the solenoid for LCLS Gun1, these fields are cancelled on average over the length of the solenoid using normal and skew quadrupole correctors. In the solenoid for Gun2, there are the same long normal and skew correctors and in addition PC quadrupole correctors are installed at both ends to locally cancel the field. Future solenoid designs should incorporate both normal and skew PC quadrupole correctors at both ends of the magnet.

### 1.1.8    Thermo-Mechanical Engineering of the LCLS Gun

This section discusses the thermal and mechanical engineering for the LCLS gun.

#### 1.1.8.1    Overview of the LCLS Gun Design

Before delving into the engineering details it's useful to first give an overview of the new gun design. Figure 19 shows a cross section of the LCLS gun for comparison with the Gun III drawing given in Figure 1 and with the LCLS specifications given in Section 1.1.4. Comparing the drawings, one observes several differences especially related to the cathode. In Gun III, the cathode plate is fully inside the vacuum envelope which requires the gun to be vented to atmosphere not only to physically change the cathode, but also to adjust the clamping of the RF seal to the body of the gun. Experience shows this is a difficult and time consuming process which significantly slows the time to recover after a cathode change due to the long exposure of the gun to air. For the case of the LCLS gun, the cathode plate is integrated with the conflat flange assembly which forms the vacuum envelope and the rear of the cathode is at atmosphere. In this approach the RF seal can be adjusted with the flanges sealed and the gun under UHV. This trivial difference turns out to be important. First it greatly reduces time it takes to change a cathode, second it allows all RF tuning to be done with the gun under UHV and third it simplifies the cathode cooling needed at the higher average power. In addition to the modifications of mode separation and RF coupling discussed in the Section 1.1.5, the new design incorporates deformation tuners and a tapered exit beam tube for wakefield mitigation.

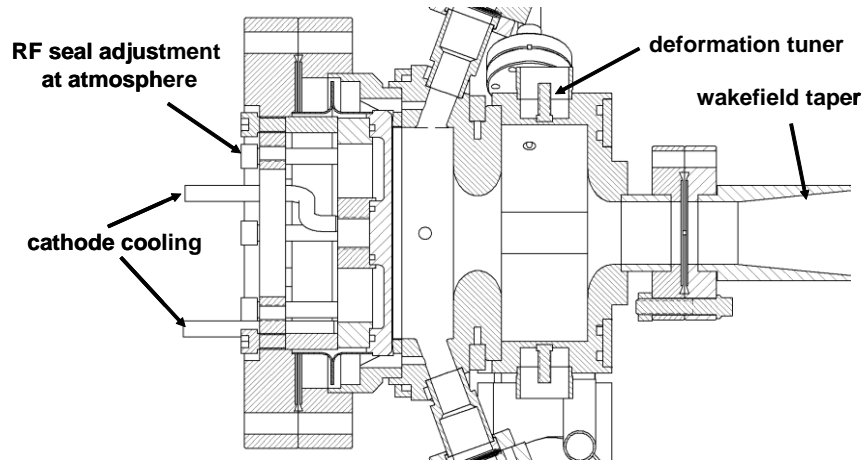

**Figure 19:** Cross-sectional drawing of the LCLS gun showing details of the cathode plate cooling, the deformation tuners and the tapered beam tube to mitigate wakefields.

### *1.1.8.2    ANSYS Simulations*

The ANSYS Finite Element Analysis package was used to analyze the effect of the RF wall losses on cavity temperature, the subsequent thermally induced deformation and the detuning of the cavity in an efficient and consistent manner. By using one program for all the simulations any problems of transferring loads were eliminated. A complete analysis cycle required six steps as outlined below:

1) The vacuum volume and the metallic structure volume were meshed with a common surface interface mesh. The analysis domain volumes were defined by Parasolids solid models exported from the Solid Edge CAD program. The common surface mesh created at this step is the key for ease of transfer of the RF wall losses onto the thermal model.
2) An Eigenmode solution of the EM fields in the vacuum volume was performed using ANSYS HF119 high frequency tetrahedral elements. The RF wall loss distribution was calculated from the eigenmode surface tangential H fields.
3) A thermal diffusion simulation of the metallic structure volume using ANSYS SOLID87 thermal tetrahedral elements was next with RF wall losses from step 2 as the thermal input and convective boundary conditions on the cooling channel surfaces as the thermal sink. The thermal flux load from the wall losses were scaled to provide a total of 4 kW average thermal load, the expected RF losses for 120 Hz 140 MV/m operation. The cooling channel convection coefficients were calculated using an Excel spreadsheet for each of the design flow conditions.
4) A thermally induced strain simulation using ANSYS SOLID92 structural tetrahedral elements loaded by the temperature field calculated in step 3 was performed next. RF surface boundary displacements and stresses occurring within the metallic structure volume were calculated at this step.

5) Next, a repeat of the eigenmode solution of step 2 using the displaced vacuum surface boundary from step 4. This step determined the cavity detuning arising from thermal distortions of the gun structure.

6) Finally, the surface displacement data from step 4 and the surface EM field data from step 2 were post-processed to calculate the cavity detuning using the Slater perturbation method [21] as a check against step 5.

This analysis cycle was repeated for designs having different cooling channel locations, cooling water flow rates, and cathode plate thicknesses until a configuration that minimized the thermally induced stress was determined. Early in this design process it was found that the radii of the z-coupling irises were too small leading to excessive heating and stress in the junction between the cavity walls and the ends of the irises. By increasing the iris radii (and changing the iris opening width and cell race racetrack dimensions to compensate) this heating and the induced stresses were reduced to acceptable levels. Figure 20 shows the quarter solid model of the ANSYS calculation of the gun's surface temperature distribution when operating at 4 kW of average power. The highest surface temperature is 36.6 °C at the RF coupler and largest temperature difference is only 10 °C.

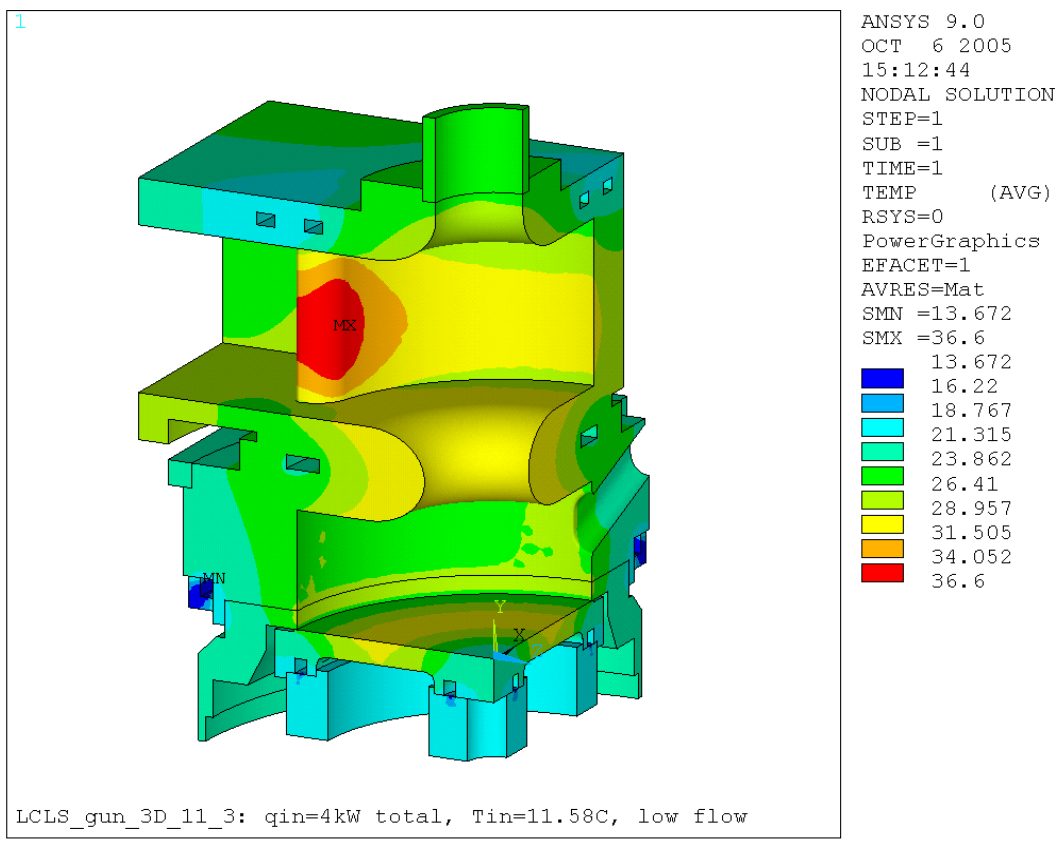

**Figure 20:** Cavity temperatures with 4kW average power dissipation and 11.6°C inlet water temperature. In this quarter model, the beam axis is vertical with the cathode at the bottom and the beam exit at the top.

An important result of the ANSYS simulations was the amount of thermal detuning of the gun that occurred at high average power and the temperature rise of the gun cooling channels over the cooling water inlet temperature. Between 0 and 4kW dissipation it was found the gun detuning to be -657 kHz which can be compensated by dropping the inlet water temperature by 13.4 °C. To reduce the amount of movement due to temperature changes of the entire gun and feed waveguide structure it was decided to set the nominal gun operating temperature to be the same as the waveguide water temperature or 35°C. A temperature controlled water cooling system is used to keep the gun on resonance over a wide range of average power operation, dropping the inlet water temperature from the nominal 35°C as required to compensate for the heating of the gun body. The system was specified with additional margin at the high and low temperature limits to allow for ±3°C thermal tuning range, equivalent to ±150 kHz, to allow more leeway in the initial gun tuning. To aid the feedback system and as an approximate indicator of resonant frequency, positions on the outside of the gun body were indentified (through the ANSYS thermal analysis) that tracked the gun average temperature, which in turn tracked the gun resonant frequency. Provisions for RTD temperature sensors were added to gun body at these locations.

### 1.1.8.3    Design, cooling and fabrication of the LCLS Gun

To allow for maximum flexibility the gun design incorporated three sets of tuning features to achieve the proper field balance and resonant frequency. Each cell had two deformable tuners at the cavity outer diameter and the center of the cathode plate was deformable. During the design phase some concern arose about our ability to tune the deformable wall tuners symmetrically and the spoiling of the quadrupole cancellation of the racetrack cavity shape of the coupler cell. To reduce the need to use these tuners, a ridge was added to the outer diameter of the cathode cell that was machined to achieve the proper tune based on cold test and bead pull field balance measurements. In the end the deformable wall tuners were never used, with the cathode plate used to tune for the design mode separation (giving the desired 1:1 field balance between the two cells in the π-mode) and the gun operating temperature adjusted to give the π-mode resonant frequency of 2856 MHz.

Based on standard SLAC klystron and RF component design practice, the gun structure was fabricated almost entirely of OFE copper due to its excellent thermal and electrical conductivity as well as the ease of joining through high temperature furnace brazing. Stainless steel (mostly the non-magnetic 304L alloy) is used in high stress locations on the outside of the gun structure and around the cathode to gun RF contact region to reinforce this critical location. Per standard design practice for beamline components, no water to vacuum joints were allowed putting significant constraints on water channel location and part design. Six cooling channels were settled on, one near the cathode ID, one near the cathode OD, one at each end of the cathode (half) cell and two at the exit of the coupling (full) cell internally plumbed as one channel. These locations allowed cooling channels that are located away from gun body brazes and that have no leak paths (except through sold metal) into the vacuum space. The cooling channels cover the entire circumference of the gun and are supplied and drained 180° apart so that flow splits into two equal parallel paths in each channel. The channels are all fed in parallel from a water manifold that distributes the temperature controlled gun water to the channels. Additional cooling channels were placed on the feed waveguide, windows, and window waveguides that were fed from another manifold with waveguide system water nominally at a constant 35°C.

Drawing on extensive experience fabricating couplers and other balanced RF structures for SLAC's x-band accelerator program it was decided to make the coupler cell (full cell) and the power splitter waveguide feeding this cell each from solid blocks of material. The waveguide structure is formed my machining a channel that forms three sides of the waveguide with a step round the top edge for brazing a cover to form the complete waveguide. While this uses a much larger block of raw material to start with the entire geometry defining the waveguide can be machined on a CNC mill in one setup so the dimensional relations between features can be very accurately controlled. This produces a very well balanced power splitter and tightly controlled phase lengths for each feed arm of the coupler cell assuring a very low dipole component in the RF fields. The accurate control of critical dimensions this technique provides lead to the lack of tuning required after final assembly of the gun. A cutaway drawing of the gun body with the dual feed is shown in Figure 21.

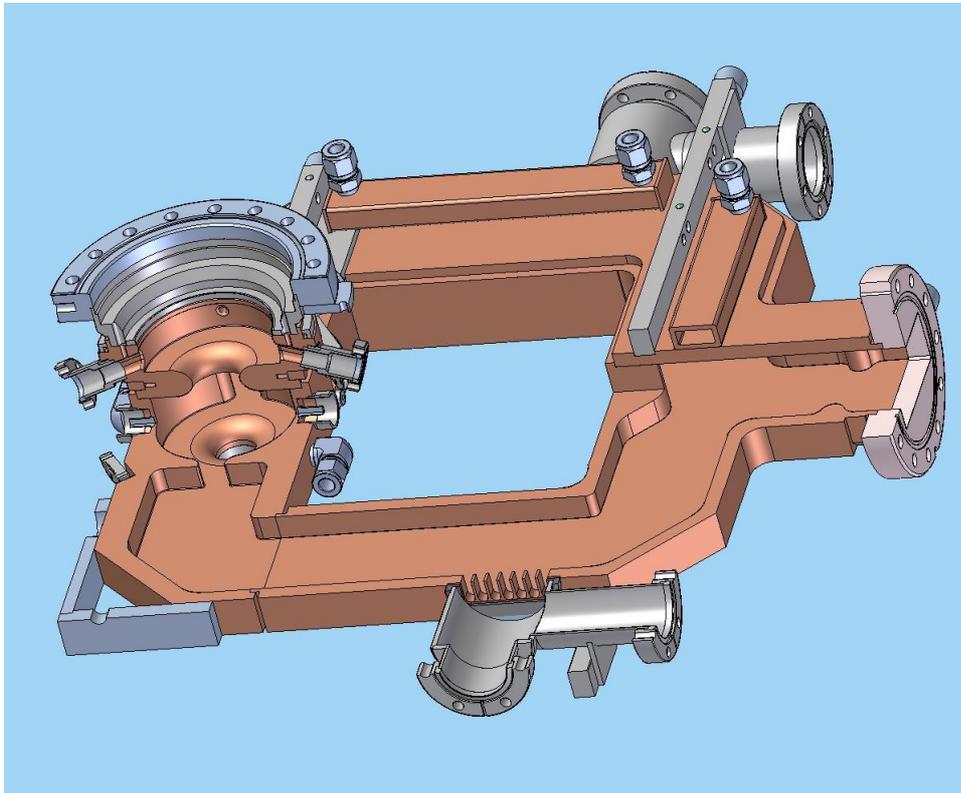

**Figure 21:** Cutaway view of the gun body showing the power splitter for feeding the balanced coupling slots in the full cell. Also shown are the silver colored stainless steel components added to the copper RF structure where additional strength was required.

### 1.1.8.4    Cathode design

The high average power requirements for the gun necessitated the inclusion of water cooling to the back side of the cathode. This significantly alters the design from that used in prior guns as it requires that cathode back side, and cooling channel covers, to be in air thus preventing the possibility of leaking water directly into the accelerator vacuum space. While complicating the design, this lead to several desirable features:

1. The RF contact surface is not the vacuum sealing surface (a feature already incorporated into the GTF and UCLA versions of Gun III).
2. The RF contact surface clamp bolts are accessible from outside the gun and can be adjusted while the gun is under vacuum. This allows for a rapid vent and cathode replacement cycle as a cathode can be removed, a new one mounted and the gun pumped down and then perform the cathode clamping and tuning.
3. The back side of the cathode can be accessed during operation to measure cathode temperature.

A cross-section of the cathode assembly is shown below in Figure 22.

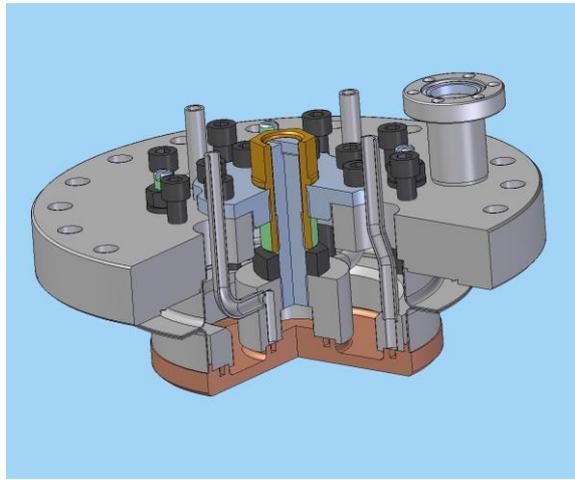

**Figure 22:** Cutaway drawing showing the details of the cathode assembly.

A cross-section of the cathode assembly and how it attaches to the gun body is shown in Figure 23. The left image is the computer solid model with a cutaway showing the cathode mounted on the gun, and the right photograph is of the gun and cathode assembly.

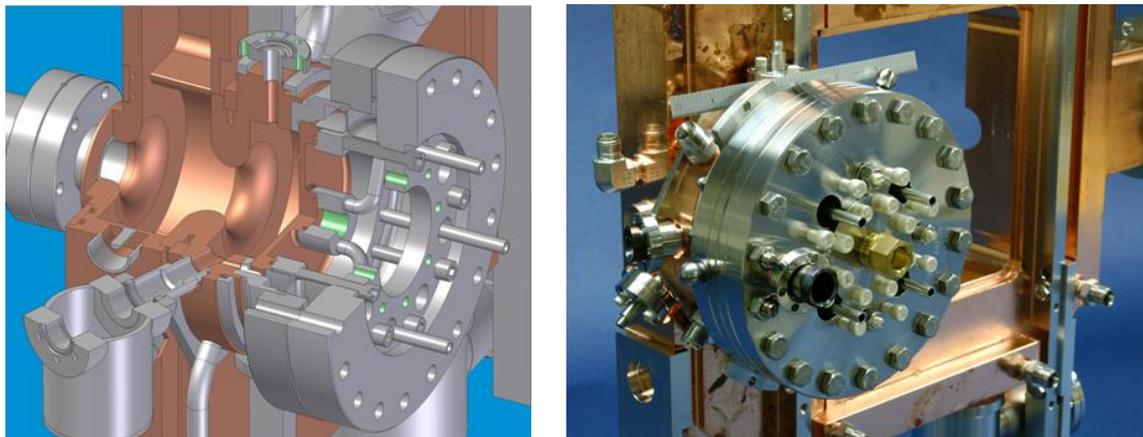

**Figure 23:** Left: Computer solid model cutaway to illustrate the interior of the gun and cathode assembly. Right: Photograph of the cathode mounted on the gun.

### *1.1.8.5 Tuner Tests*

The tuner design selected for the gun was based on a deformable wall tuner used for x-band klystron cavity and accelerator cell tuning, scaled to a larger size more appropriate for S-band. To test the new tuner design, a pillbox cold test cavity was fabricated with a series of eight tuning features around the circumference. The tuning features had two different diameters and three different wall thicknesses to study the interaction of wall thickness and diameter. A threaded stud was brazed into the center of each tuner to allow tuning either direction by pushing or pulling the stud. Each tuner was tested by tuning both inward and outwards to destruction (in all cases the stud pulled out with no rupturing of the tuner wall). The best tuner geometry was found to easily achieve +/- 1.5 MHz of tuning.

### *1.1.8.6 Cathode Seal Tests*

The unique design for the cathode mount has one disadvantage, the clamp screws for loading the RF seal contact introduce a moment on the conflat flange that unloads the vacuum seal possibly leading to a leak. A test fixture was made to test the clamp screw load that could be applied before the vacuum seal is compromised. Testing showed at least a factor of three safety margin between the nominal clamp screw force (sufficient to seat the cathode RF contact) and the clamp screw force that unloaded the conflat seal sufficient to start leaking. Subsequent installation of multiple cathodes in two guns has shown no problems with vacuum sealing of the cathode conflat flange.

### *1.1.8.7 Integration of the LCLS Gun with the Emittance Compensation Solenoid*

The final assembly of the LCLS gun and emittance compensation solenoid is mounted on a single strong back plate as shown in the photographs of Figure 24. The system can be transported and installed without disturbing the ultra high vacuum of the gun. Dual RF vacuum windows isolate the gun vacuum from the klystron waveguide vacuum in order to halve the power load on the windows. However operational experience has shown that this is not necessary and in the future a single RF window will be used. The mechanical alignment between the gun and solenoid is performed by moving the solenoid while the gun remains stationary on its mount.

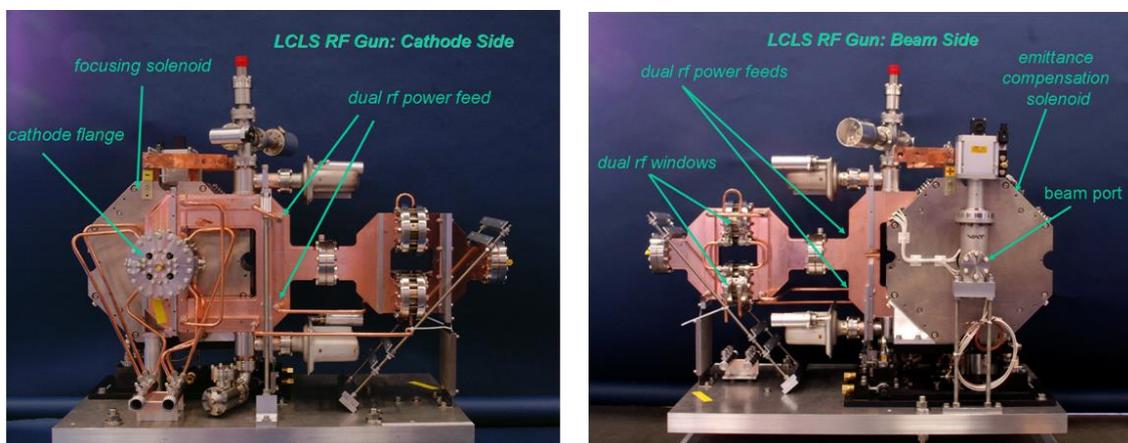

**Figure 24:** Photographs of the LCLS gun and solenoid assembly. Left: View from the cathode side. Right: View of the beam exit side.

### 1.1.9   The Cold and Hot RF Testing

#### *1.1.9.1    RF Cold Tests and Tuning the Gun*

As described in Section 1.1.8.3, a tuning ridge 3.8 mm wide and 1.0 mm tall at the cavity OD placed longitudinally 3.8 mm from the cathode plate (in z) was incorporated to allow tuning of the gun resonance frequency and field balance. The nominal final height in the cavity radial direction with the gun properly tuned should be 0.5 mm tall. The field balance was measured using the beam drop method similar to the data shown in Figure 3. Three successive iterations were made to machine off this ridge, based upon the bead drop and resonant frequency measurements until the desired π-mode frequency, mode separation and field balance were obtained. When this was done, the final ridge dimension was found to be identical to the RF design value. Although incorporated into the gun design, the deformation tuners were not needed to achieve the final tune parameters. The above measurements were made with the gun parts clamped together which then could be brazed and welded into the final gun assembly.

After the final braze a cathode was installed and the tuning curve consisting of the field balance as a function of the mode separation frequency was measured by slightly deflecting the cathode plate with the differential tuning screw at center of the cathode and using the bead drop to determine the field balance. The results are shown in Figure 25. Additional tests were done to confirm the gun could be tuned to the correct frequency, Q and field balance based simply upon the mode separation frequency by repeatedly removing and re-installing the cathode. These tests are essential to establish the procedure for in situ tuning of the gun on the beamline when the cathode is changed and a bead drop cannot be performed. The final RF characteristics are compared with the design parameters in Table 4.

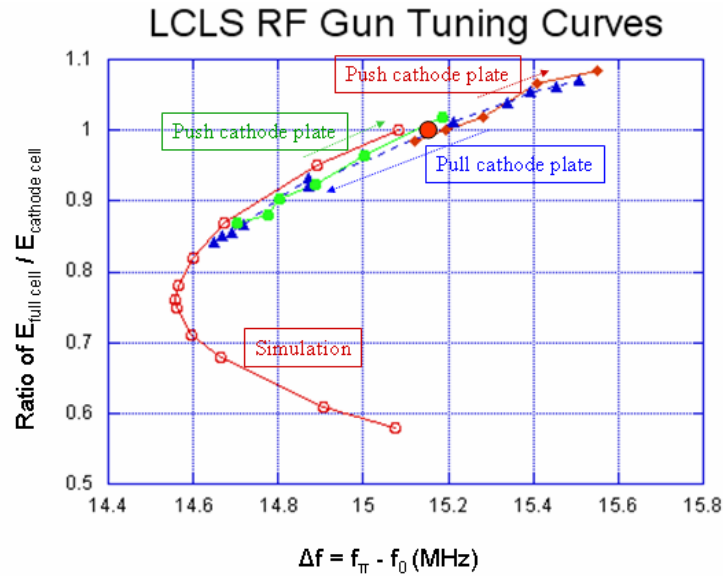

**Figure 25:**  Field balance as a function of mode separation frequency [22].

**Table 4:** The design and measured parameters of the LCLS Gun

| RF Parameters | Design | Measured |
|---|---|---|
| $f\pi$ (GHz) | 2.855987 | 2.855999 |
| Q0 | 13960 | 13900 |
| β | 2.1 | 2.03 |
| Mode Sep. Δf (MHz) | 15 | 15.17 |
| Field Balance | 1 | 1 |

### 1.1.9.2    RF Hot Tests of the LCLS Guns

With the completion of the cold tests, the gun was assembled with the emittance compensation solenoid onto a common base plate and installed in a radiation shielded vault in the SLAC Klystron Lab.  Here it was vacuum baked to approximately 160 °C using electrical heater tape.  After the bake and returning to room temperature, the gun vacuum pressure was in the mid-$10^{-10}$ torr range.  The baked gun was then RF processed to a peak cathode field of 120 MV/m, 2 microsecond long RF pulse and 120 Hz repetition rate [22] as shown in Figure 26.

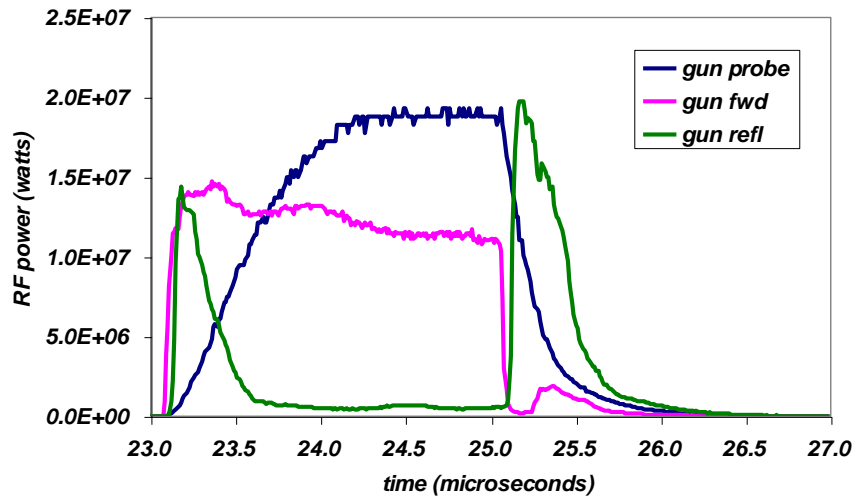

**Figure 26:** RF waveforms measured during high power testing of the LCLS gun [22].

As described earlier, the RF probes experienced excessive RF heating during operation at 120 Hz.  These probes had an RF coupling of -55 to -60 dB to the cell fields and consisted of stainless steel rods attached to SMA-style electrical connectors on mini-conflat flanges.  Due to this heating, it was decided to limit the operation of this first LCLS gun, Gun1, to 30 Hz repetition rate until new probes could be designed and tested in the second LCLS gun, Gun2.

The new probes design utilized copper-plated stainless steel rods to reduce the electrical resistance, a lower RF coupling of -76 to -80 dB, and were mounted using more robust type-N electrical feedthroughs.  These were installed in Gun2 and successfully operated to 120 Hz and 125 MV/m peak cathode field with an average dissipated power of 2 kilowatts.  The high average power allowed the cathode peak field to be determined by both the forward RF power and to be derived from the temperature rise of the gun cooling water. The correlation of the peak cathode field using these two techniques is plotted in Figure 27.

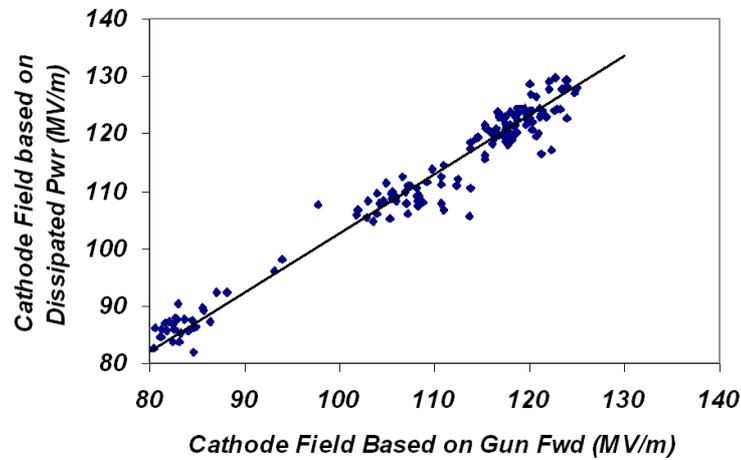

**Figure 27:** The cathode peak field as determined from the dissipated power correlated with the forward RF power during RF conditioning of Gun2.

After the new probes were successfully demonstrated on Gun2, a similar set were installed in Gun1 which had been operating nearly continuously for a year in support of LCLS injector and linac commissioning. The photographs in Figure 28 show the rear of Gun1 with the old (left) and new (right) probes. The change from SMA-type to N-type connector can be seen. With this upgrade the operational limit of 30 Hz is removed, and full power operation with beam at 120 Hz is expected during the ongoing 2008 commissioning run.

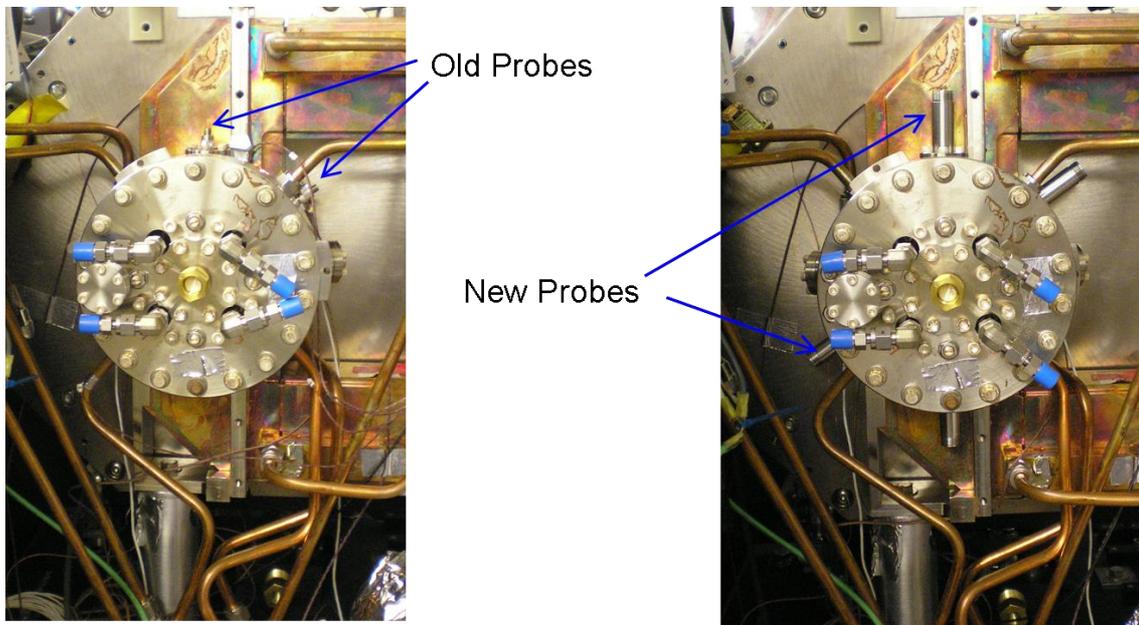

**Figure 28:** Rear views of Gun1 installed in the LCLS injector. Left: Gun1 with the original RF probes. Right: Gun1 with the upgraded probes capable of 120 Hz full power operation.

### 1.1.10 Summary and Conclusions

The complete tested gun and solenoid assembly of LCLS Gun1 was installed on the LCLS injector in March 2007 and immediately began 30 Hz beam operations. The RF probes were upgraded with the new probes in April 2008, and the gun is now capable of full power operation at 120 Hz. The only other operational difficulty has been the cathode quantum efficiency which was initially $4 \times 10^{-6}$ or 15-times lower than the $6 \times 10^{-5}$ design specification. This was later increased to $4.1 \times 10^{-5}$ by cleaning the cathode with the UV drive laser. However this cleaning resulted in damage to the cathode surface and had to be done repeatedly due to a recent vacuum leak in the beamline. As a result, a new cathode prepared using a different preparation process was installed in July 2008. The gun vacuum recovered within a day and resumed operations showed the new cathode had a much improved quantum efficiency of $5 \times 10^{-5}$ [23]. Processes for improving the quantum efficiency and cathode emission uniformity are the topics of ongoing studies.

LCLS Gun1 has operated reliably since April 2007 with excellent beam quality which continues to improve with operational experience. Figure 29 illustrates the projected emittance as a function of the bunch charge measured at 135 MeV using the quadrupole scan technique with an optical transition radiation screen. The values and error bars are for multiple measurements made on different days. Further information on the beam measurements and other details for the LCLS injector can be found in [2, 23].

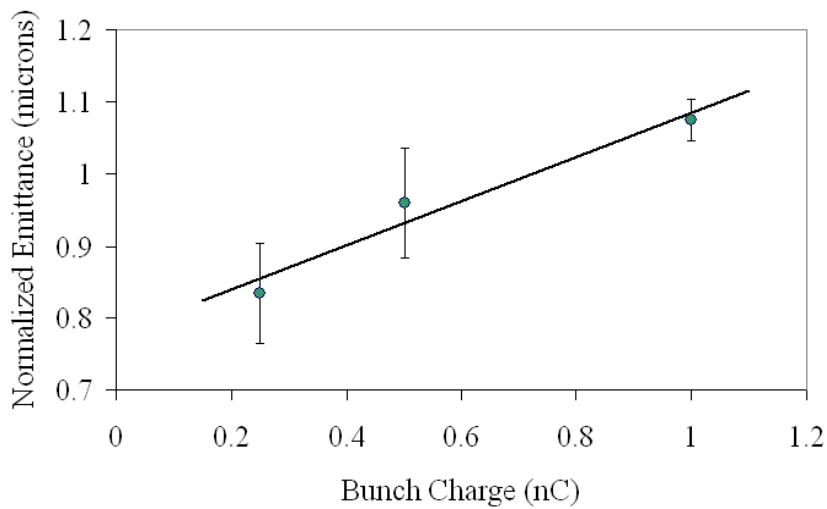

**Figure 29:** The projected rms emittance for bunch charges of 0.25, 0.50 and 1.0 nC. The emittances are the average of the x- and y-plane emittances and were measured at 135 MeV.

The photocathode RF gun described in this paper has achieved the stringent requirements needed for the operation of the LCLS x-ray free electron laser. This success was the result of experimental studies of a prototype gun to understand its limitations and to determine the modifications necessary to achieve the LCLS requirements. These modifications and operational experience were then applied in the engineering and construction of a new gun. The result is the LCLS gun which is an enabling technology for the new era of 4th generation light sources.

### 1.1.11 Acknowledgements

We wish to thank all the people and SLAC technical groups that contributed to the success of this RF gun design: ACD for the countless RF simulations, Klystron and Microwave Department for fabricating and testing the gun, Magnetic Measurements group for solenoid characterization, MFD Vacuum group for final assembly of the gun, the SLAC Precision Alignment group and the LCLS project for providing the funds to execute such a successful design. We also thank Herman Winick for providing the history of the GTF at SLAC.